\documentclass[11pt,preprint,tighten]{aastex}
\usepackage{graphicx}
\usepackage{times}
\usepackage{rotating,lscape}
\usepackage{epsfig}
\usepackage{graphicx}
\usepackage{epstopdf}

\usepackage{xcolor}
\usepackage{booktabs}
\usepackage{amsmath}
\usepackage{color,soul}
\usepackage[flushleft]{threeparttable}

\def\chandra{{\it Chandra}~}
\def\xmm{{\it XMM-Newton}~}

\def\chandran{{\it Chandra}}
\def\xmmn{{\it XMM-Newton}}

\def\lya{\ifmmode {\rm Ly}\alpha~ \else Ly$\alpha$~\fi}
\def\lyan{\ifmmode {\rm Ly}\alpha \else Ly$\alpha$\fi}
\def\lyb{\ifmmode {\rm Ly}\beta~ \else Ly$\beta$~\fi}
\def\lyg{\ifmmode {\rm Ly}\gamma~ \else Ly$\gamma$~\fi}
\def\civ{\ifmmode {\rm C}\,{\sc iv}~ \else C\,{\sc iv}~\fi}
\def\civn{\ifmmode {\rm C}\,{\sc iv}~ \else C\,{\sc iv}\fi}
\def\cvi{\ifmmode {\rm C}\,{\sc vi}~ \else C\,{\sc vi}~\fi}
\def\cvin{\ifmmode {\rm C}\,{\sc vi} \else C\,{\sc vi}\fi}

\def\ovi{{{\rm O}\,{\sc vi}~}}

\def\ovii{{{\rm O}\,{\sc vii}~}}
\def\oviii{{{\rm O}\,{\sc viii}~}}

\def\neviii{{{\rm Ne}\,{\sc viii}~}}

\def\neix{{{\rm Ne}\,{\sc ix}~}}

\def\gax{${_>\atop^{\sim}}$}

 \def\pks{{\it PKS 0405-123}~}
\def\pksn{{\it PKS 0405-123}}

\begin{document}
\title{Probing the Warm-Hot Circumgalactic Medium with broad \ovi and X-rays}

\author{Smita Mathur\altaffilmark{1,2}, Anjali Gupta \altaffilmark{1,3}, Sanskriti Das\altaffilmark{1}, Yair Krongold\altaffilmark{4} and Fabrizio Nicastro\altaffilmark{5} }
\altaffiltext{1}{Astronomy Department, The Ohio State University, Columbus,
  OH 43210, USA} 
\altaffiltext{2}{Center for Cosmology and
  Astro-Particle Physics, The Ohio State University, Columbus, OH 43210}
\altaffiltext{3}{Columbus State Community College, Columbus, OH, USA}
\altaffiltext{4}{Instituto de Astronomia, Universidad Nacional Autonoma de Mexico, 04510 Mexico City, Mexico}
\altaffiltext{5}{Observatorio Astronomico di Roma - INAF, Via di Frascati 33, 1-00040 Monte Porzio Catone, RM, Italy}

\begin{abstract}
Most of the baryonic mass in the circumgalactic medium (CGM) of a spiral galaxy is believed to be warm-hot, with temperature around $10^6$K. The narrow \ovi absorption lines probe a somewhat cooler component at $\log \rm T(K)= 5.5$, but broad \ovi absorbers have the potential to probe the hotter CGM. Here we present 376 ks \chandra LETG observations of a carefully selected galaxy in which the presence of broad \ovi together with the non-detection of \lya was indicative of warm-hot gas. The strongest line expected to be present at $\approx 10^6$K is \ovii $\lambda 21.602$. There is a hint of an absorption line at the redshifted wavelength, but the line  is not detected with better than $2\sigma$ significance.  A physical model, taking into account strengths of several other lines, provides better constraints. Our best-fit absorber model has $\log \rm T(K) =6.3\pm 0.2$ and $\log \rm N_{H} (cm^{-2})=20.7^{+0.3}_{-0.5}$. These parameters are consistent with the warm-hot plasma model based on UV observations; other \ovi models of cooler gas phases are ruled out at better than $99$\% confidence. Thus we have suggestive, but not conclusive evidence for the broad \ovi absorber probing the warm-hot gas from the shallow observations of this pilot program. About 800ks of \xmm observations will detect the expected absorption lines of \ovii and \oviii unequivocally. Future missions like XRISM, \textit{Arcus} and \textit{Athena} will revolutionize the CGM science.
\end{abstract}

\section{Introduction}

In recent years, the circumgalactic medium (CGM) of spiral galaxies has
attracted a lot of scientific attention (see the review by \citet{Tumlinson2017}). The CGM is the gaseous medium
surrounding the stellar disk of a galaxy, extended out to its virial
radius (or perhaps twice the virial radius \citep{Oppenheimer2016}. It serves as a gas reservoir with accretion from the
intergalactic medium (IGM) and outflows from the stellar disk
(star-formation and/or AGN induced). Some of this material may recycle
back into the disk of the galaxy, while some may stay in the diffuse
CGM. The CGM may help sustain and regulate star-formation and may harbor
the largest baryon and metal reservoir of galaxies. As such, it is a very important
component of a galaxy, but we are just about beginning to understand
it. There have been  significant theoretical developments in recent years 
(e.g. \citep{Oppenheimer2016}), giving us insight into
the physics of the CGM and also attesting to the importance of the CGM
in the overall formation and evolution of galaxies.

The CGM is believed to be multi-phase, with most of its baryonic mass at
temperatures of $10^5$--$10^7$K, historically referred to as ``warm-hot" \citep{Cen1999}. The cooler range of this, the warm CGM at $\approx 10^5$--$10^{5.5}$ K is probed by
narrow UV absorption lines of Li-like metals; we will refer to this range as "warm" in the rest of the paper.  
The  \ovi absorption lines in the UV are among the best probes of the warm CGM (\citep{Tumlinson2017} and references therein).

At higher temperatures ($\approx
10^6$--$10^7$K) metals become more highly ionized, He-like and H-like,
observable only through X-ray absorption and emission lines; we will refer to this temperature range as ``warm-hot" in the following sections. 
Because of our special vantage point, the warm-hot CGM has been best studied for 
our own galaxy, the Milky Way, both in X-ray absorption \citep{Nicastro2002,Wang2005,Williams2005,Fang2006,Bregman2007,Gupta2012,Gupta2014,Gupta2017,Das2019a} and emission (\citep{Snowden2000,Smith2007,Galeazzi2007,Henley2007,Henley2008,Gupta2009,Das2019c}). 
X-ray observations of the CGM of external galaxies have had only limited
success (e.g. \citep{Anderson2016} and references
therein). Most of the observations were for {\it massive} galaxies; they were detected in X-ray emission, but
the physical state, extent, temperature and density distribution
profiles, and mass content of the CGM of these galaxies are poorly
constrained. The only external L$^{\star}$ galaxy in which the warm-hot CGM is detected in X-ray emission is NGC\,3221 \citep{Das2019b,Das2020a}.

Blind searches of the warm-hot CGM of external galaxies with X-ray absorption spectroscopy are difficult for the lack of X-ray bright background quasars. Even with large exposure times, the resulting spectra have low signal-to-noise ratio (S/N), making it hard to assess the significance of  the intervening absorption lines. With redshifts known {\it a priori} we can place higher confidence in
the lines with lower detection thresholds, because we no longer have to
pay the statistical penalty for searching a large number of resolution
elements over the entire spectrum. This can be achieved by using ``UV markers''; we can look for X-ray lines at redshifts defined by the UV absorption lines. Apart from the rare \neviii lines (\citep{Frank2018} and references therein), \ovi are the highest ionization lines probing the warm gas, making them suitable anchors for X-ray searches. Moreover, oxygen is also one of the most abundant elements, making \ovi absorbers promising probes of the warm gas. Even better anchors would be the broad \ovi absorbers; in these systems the \ovi lines are thermally broadened to $b>25 \rm km~s^{-1}$ indicative of warm-hot gas at $T \geq 10^{5.7}~$K. The line broadening, however, could be because of non-thermal effects, such as turbulence, so not all broad \ovi absorbers probe the warm-hot gas. 

%Stocke et al. (2014, ApJ, 791, 128) and Savage et al. (2014, ApJS, 212,
%8, hereafter SA14; 2010, ApJ, 719, 1526, hereafter SA10) 

 \cite{Savage2010} (hereafter SA10), \cite{Stocke2014} and \cite{Savage2014} (hereafter SA14) conducted a
high S/N survey of 14 bright QSOs using HST/COS to discover warm
($10^{5.5}$~K) intervening absorbers. In the spectrum of \pksn, SA10 reported the discovery of a symmetric, broad \ovi absorber (b=$52\pm2~$km~s$^{-1}$) at
$z\approx0.167$. This is the first, and the only broad \ovi
  intervening absorption system with no associated \lyan, making this a
unique laboratory for CGM studies.  The line symmetry, the $b$-value and the absence of
associated \lya absorption strongly suggest that it is produced in
collisionally ionized warm-hot gas with $\log T (K) \approx 6.41$ and $\rm \log
N_{H} (cm^{-2}) \approx 20.5$ (SA10; SA14), and it likely originates in the CGM
associated with a galaxy at z=0.1668. It is a luminous spiral
$4.4L^{\ast}~$ galaxy within $\Delta v < 100~km~s^{-1}$ of the absorber
redshift and with an impact parameter of 116 kpc.  SA10 do not rule out the
possibility that the \ovi absorption is produced in cooler gas with
$\log T (K)=5.8$ (see table \ref{tab:Allmodels} for all the possible solutions from SA10). The line broadening in that case could be because of
non-thermal effects and the lack of \lya may then be
attributed to high oxygen abundance.

We observed this unique and interesting system with \chandra to look for the presence of the warm-hot gas from the CGM of the galaxy at $z=0.167$. The observations, data analysis, and spectral modeling are discussed in \S2. We show that there is suggestive, but not conclusive evidence for the warm-hot gas in the CGM of the galaxy. In \S3 we show that 800 ks of \xmm observations will unequivocally detect the warm-hot gas, and {\textit Athena} will require only 40 ks, opening a new window into the CGM science. Our conclusions are presented in \S4. 

\section{\chandra Observations, Analysis, and Results}

Our 385 ks observations of the background quasar \pks with \chandra LETG/HRC-S (PI: Gupta) took place in October and November of 2018. The actual exposure time was 376.36 ks and the effective LETG exposure time was 376.1 ks. We reprocessed the \chandra data  with \chandra interactive analysis of observations (CIAO) software and extracted the grating spectra following the standard procedure in CIAO v4.10\footnote{https://cxc.harvard.edu/ciao/}. Since HRC-S does not have clear order sorting, we included grating orders 1--6 in our analysis and combined the plus and minus orders. The spectrum showed practically no data beyond 3 keV, so we analyzed the spectrum in the 0.3--3 keV range. 
 We binned the spectrum by four, to the LETG resolution; the count rate was found to be $0.1$ counts/s. All the spectral analysis was performed with XSPEC v12.10.1\footnote{https://heasarc.gsfc.nasa.gov/xanadu/xspec/}. In all spectral models absorption by the Galactic column density of $\rm N_{H}(Gal)=3.54 \times 10^{20}$ cm$^{-2}$ \citep{BenBekhti2016} was included and held fixed. The errors are quoted at $90$\% confidence, unless noted otherwise.
 
 \subsection{Model-independent analysis}

\begin{figure}
\begin{center}
\includegraphics[width=5cm]{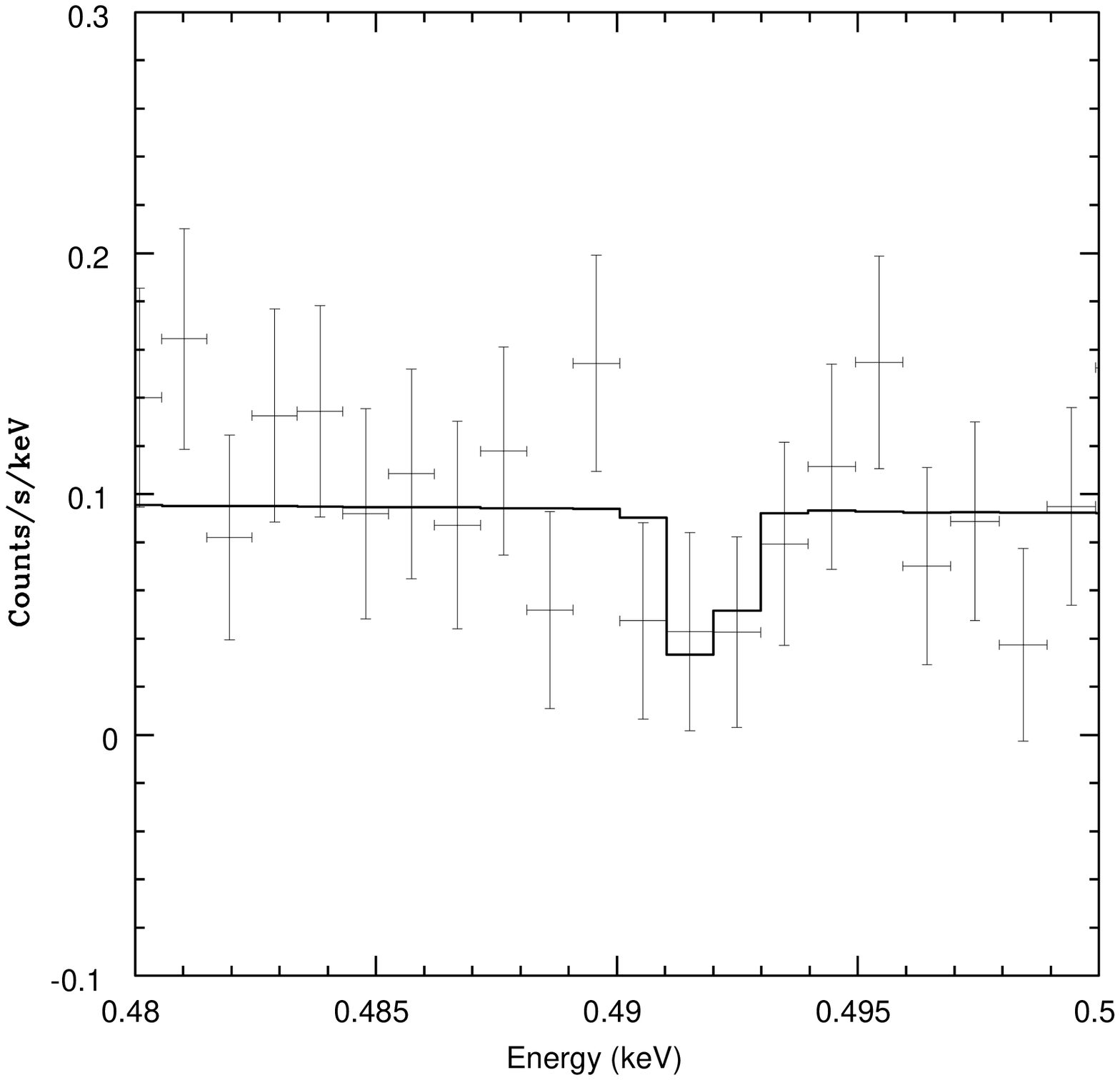}
\includegraphics[width=5cm]{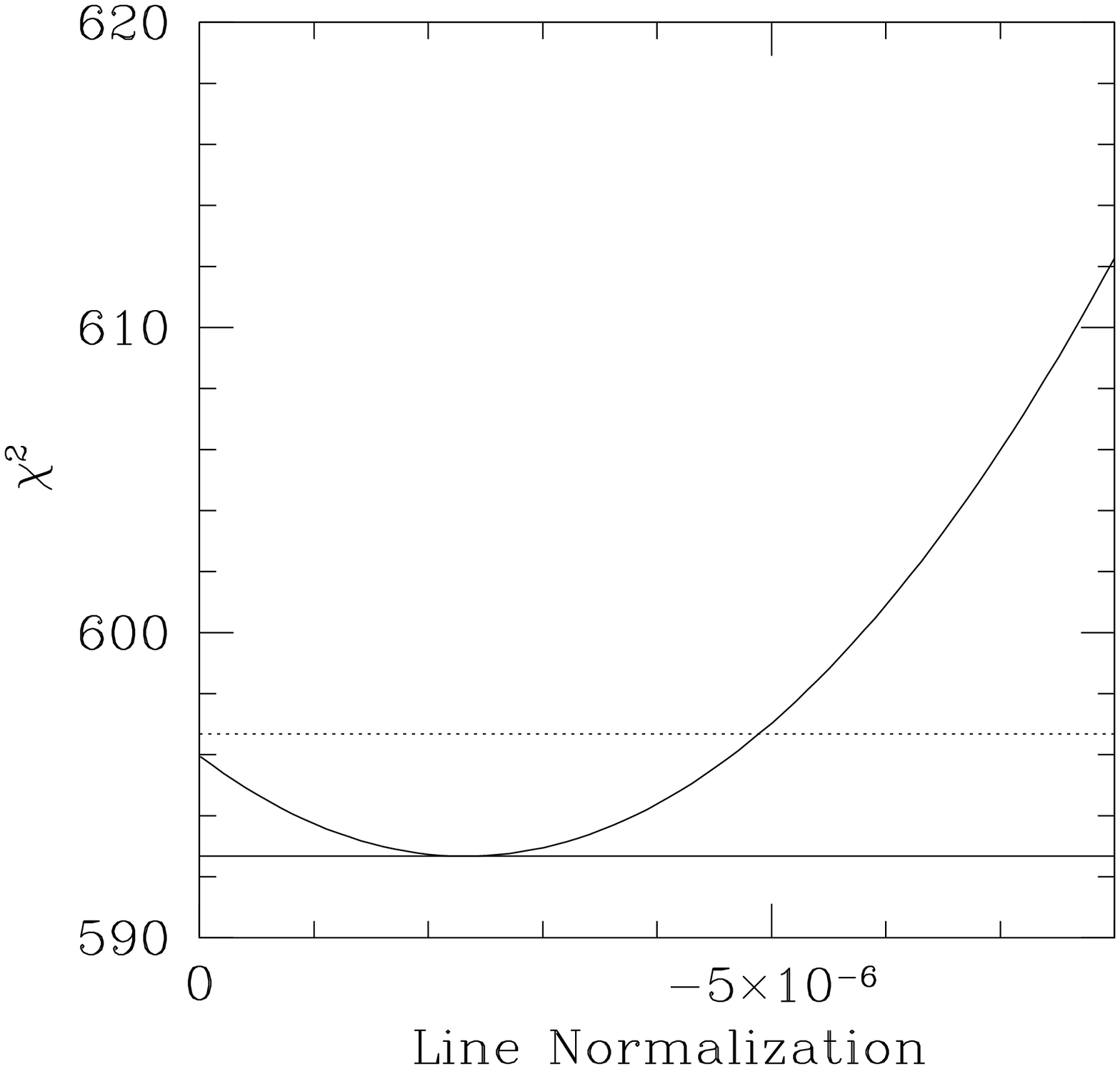}
\end{center}
%\vspace*{-2cm}
\caption{Left: \chandra LETG spectrum fitted with an absorbed power-law continuum and a Gaussian absorption line at 0.492 keV (25.21\AA; \ovii $\lambda 21.602$ at z=0.167). Right: The $\chi^2$ curve of the Gaussian line normalization (in the units of photons cm$^{-2}$ s$^{-1}$ keV$^{-1}$). The solid horizontal line is for the best-fit $\chi^2=592.67$ and the dashed horizontal line shows $\Delta \chi^2=4$ corresponding to the $2\sigma$ significance. Thus the line is detected at just below $2\sigma$ significance.} 
\label{fig:linefit}
\end{figure}

We fitted the LETG spectrum with a power-law continuum and looked for the presence of the \ovii absorption line at the redshift of the intervening galaxy. 
As noted above, the advantage of knowing the absorber redshift {\it a priori} is that we know the wavelength/energy of the expected lines; we therefore do not need to search for the lines over the entire spectrum. The strongest line expected from the warm-hot CGM is \ovii $\lambda 21.602$; this gets redshifted to $25.21$\AA\ at $z=0.167$. Therefore we looked for a line signature close to this wavelength (at E$=0.492$ keV). A line-like feature was present at the expected energy, so we added a Gaussian absorption line to the continuum model and fitted again. The line width was kept well below the instrumental resolution, to $\sigma=0.1$ eV; thus the observed line width is due to the grating resolution. The best-fit parameters were as follows. The power-law $\Gamma=1.94\pm 0.04$ and normalization $=6.9\pm 0.1\times 10^{-4}$ photons cm$^{-2}$ s$^{-1}$ keV$^{-1}$ and the line normalization $=-2.3\pm 2.1\times 10^{-6}$ photons cm$^-2$ s$^{-1}$ ($90$\% confidence) (see figure \ref{fig:linefit}). 

The best-fit absorption line EW$=1.1\pm0.6$ eV ($1\sigma$ error bars). The hard limit of normalization $=0$ is reached at  $\Delta \chi^2=3.26$, below the $2\sigma$ confidence of $\Delta \chi^2= 4$ (corresponding to $95.45$\% confidence). 
To evaluate the line detection significance further, we generated the $\chi^2$ contour of the line normalization, as shown in figure \ref{fig:linefit}, right. The horizontal line marks the $2\sigma$ confidence interval. Thus we see that the line is detected at just below $2\sigma$ significance. 

\subsection{Modeling}

In the above analysis, our focus was on the \ovii line, which is expected to be the strongest at the warm-hot temperature. However, there is additional information in detection/non-detection of other H-like and He-like ions. Together, they can inform us about the temperature and column density of the CGM probed. With this in mind, we fitted the LETG spectrum with our hybrid photo- and collisional-ionization model hPHASE \citep{Krongold2003}. The free parameters of the model are plasma temperature T, column density $\rm N_{H}$, and the (non-thermal) microturbulent velocity $v$. Since the warm-hot CGM is expected to be collisionally ionized, we kept the model photo-ionization parameter frozen to its lowest allowed value of $\log U= -4$, with negligible effect on the observed line ratios in the warm-hot plasma. The best-fit model parameters are $\log \rm T (K)=6.3\pm 0.2$, $\log \rm N_{H} (cm^{-2})=20.7^{+0.3}_{-0.5}$, and $v=12^{+124}_{-12}$ km s$^{-1}$ (Table \ref{tab:Allmodels}). The microturbulent velocity is not constrained, but shows that the lines are not resolved. On the other hand, the temperature and column density are very well constrained and the measured temperature is consistent with the observed b-value of the broad \ovi absorption lines. This shows the power of model fitting using many lines of many ions. We note that the best-fit model \ovii line EW$=34.43$ m\AA\ is similar to the EW$=41\pm22$m\AA\ obtained by the Gaussian line fitting.
In figure \ref{fig:contour} we present the $\log \rm T$ -- $\log \rm N_{H}$ contour plot from the spectral fit. The "+" symbol refers to the best-fit values and the solid, dotted and dot-dash lines correspond to $68.3\%$\, ($1\sigma$), $90\%$\, ($1.64\sigma$), and $99\%$\, ($2.58\sigma$) confidence contours respectively. We see that the parameters are well-constrained at $68.3\%$ and $90\%$, but not at $99\%$.

\section{Discussion}

Our hPHASE model contains more than 100 lines of multiple ionization states of ten elements (C, N, O, Ne, Mg, Si, S, Ar, Ca and Fe). Consistency of model predictions with data for all the lines provides a powerful tool to constrain the physical parameters of the absorbing warm-hot plasma. The upper limit on the temperature is driven by the \oviii line strength/weakness, and the lower limits by the strengths/weakness of lower-ionization lines, even though the lines are not individually detected. The column density is constrained by the line strength at a given temperature. This is what we see with the good constraints on $\rm T$ and $\rm N_{\rm H}$ in the above analysis. 

\begin{figure}
\begin{center}
\includegraphics[width=7cm,angle=-90]{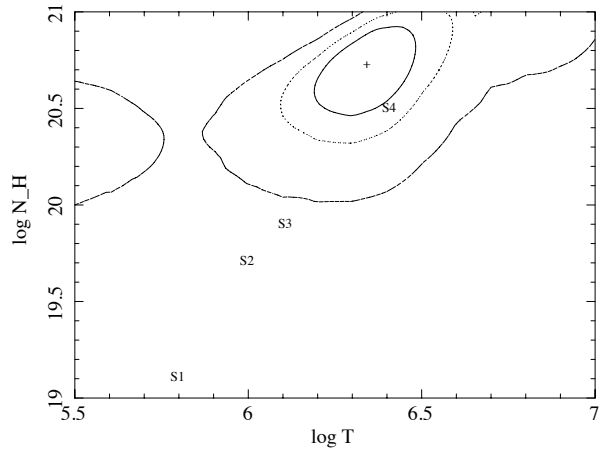}
\end{center}
\caption{The contour plot showing the temperature--column density plane. The "+" denotes the best-fit values, and the sold, dotted and dot-dash lines correspond to $68.3\%$, $90\%$, and $99\%$ confidence contours respectively. S1, S2, S3, and S4 refer to the four model parameters in Savage et al. 2010. The best-fit model is consistent with S4, and S1, S2, S3 are ruled out at better than 99\% confidence. }
\label{fig:contour}
\end{figure}

As noted in \S 1, SA10 argue that the broad \ovi absorber in the spiral galaxy at $z=0.167$ likely traces warm-hot gas. However, they could not rule out other models with cooler temperatures (SA10; their Table 5). We refer to their four models as S1, S2, S3, and S4; these are presented in table \ref{tab:Allmodels} and figure \ref{fig:contour}. We see that our best-fit model is closest with  model S4 ($\log T(K) =6.41$) and is consistent with S4 with $90$\% confidence. S1, S2, and S3 are ruled out at larger than $99$\% confidence. Once again, the consistency of the data on several lines with the model predictions allows for distinguishing among the models. For example, the best-fit model equivalent width of the \oviii K$\alpha$ line is EW$=19.68$ m\AA, similar to what is predicted by S4: EW$=17.19$ m\AA. The predicted line EW for S1, however, is less than 1m\AA. Similarly, the best-fit model predicted EW for the \neix $\lambda 13.447$ is EW$=13.74$ m\AA, similar to EW$=10.91$ for S4. But the line EW is predicted to be less than a m\AA\ for S1.

\begin{figure}
\begin{center}
\includegraphics[scale=0.4]{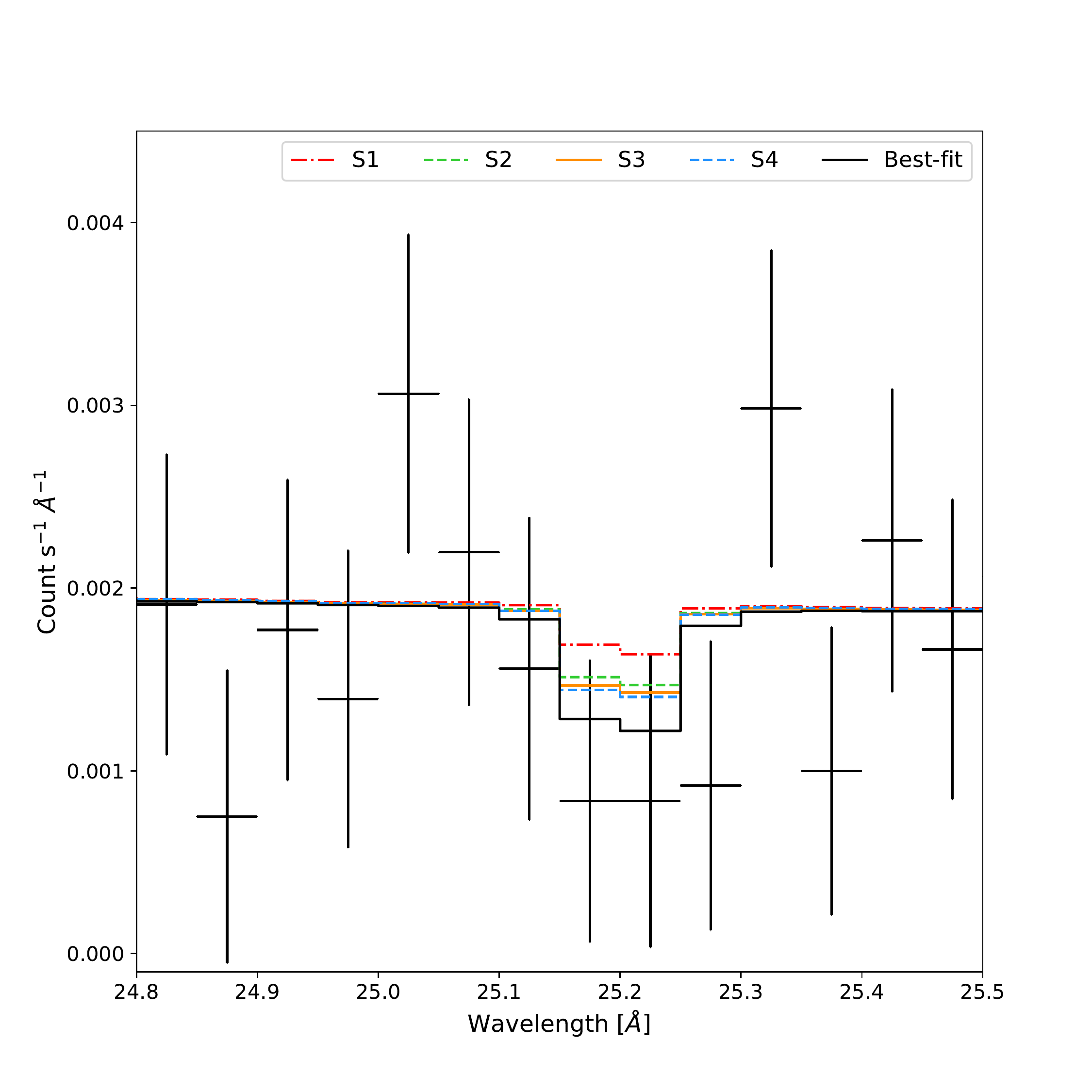}
\end{center}
\caption{ The spectrum around the \ovii line with best-fit (black), S1(dot-dashed red), S2 (dashed green), S3 (solid orange) and S4 (dashed cyan) models. The S4 model is closest to the data and the best-fit model. The S1, S2 and S3 models are close to each other and the \ovii line strength is also consistent with them.}
\label{fig:specmodels}
\end{figure}

The above results, however, are model-dependent. 
In our models we have assumed that the warm-hot CGM is in collisional ionization equilibrium; while this is a reasonable assumption for an L$^{\star}$ galaxy like the Milky Way, photoionization may play a non-negligible role in a  4L$^{\star}$ galaxy that we probe here. However, as noted by SA10, for the photoionization model to be valid, the non-thermal microturbulent velocity needs to be associated with a symmetric flow to explain the symmetric line profile. It would have been nice to obtain model-independent results, but 
 the weak detection of the \ovii line makes the model distinction much more difficult based on the observed line strength. This can be seen in figure \ref{fig:specmodels} where the black line is for the best-fit model and models S1, S2, S3 and S4 are shown by red, green, orange, and cyan lines respectively. We again see that the best-fit model is closest to S4, but S1, S2, and S3 cannot be ruled out at better than $2\sigma$.  This shows that we should be cautious with the model-dependent results, but it also shows the power of model fitting. In figure \ref{fig:specmodels} we also see that the model predictions for the \ovii line are close for S2, S3 and S4, even though they correspond to quite different temperatures and column densities as shown in figure \ref{fig:contour}. Once again, this distinction comes because the model includes many lines of many elements. 

\subsection{The power of \xmm and future missions}

\begin{figure}
\begin{center}
\includegraphics[scale=0.75]{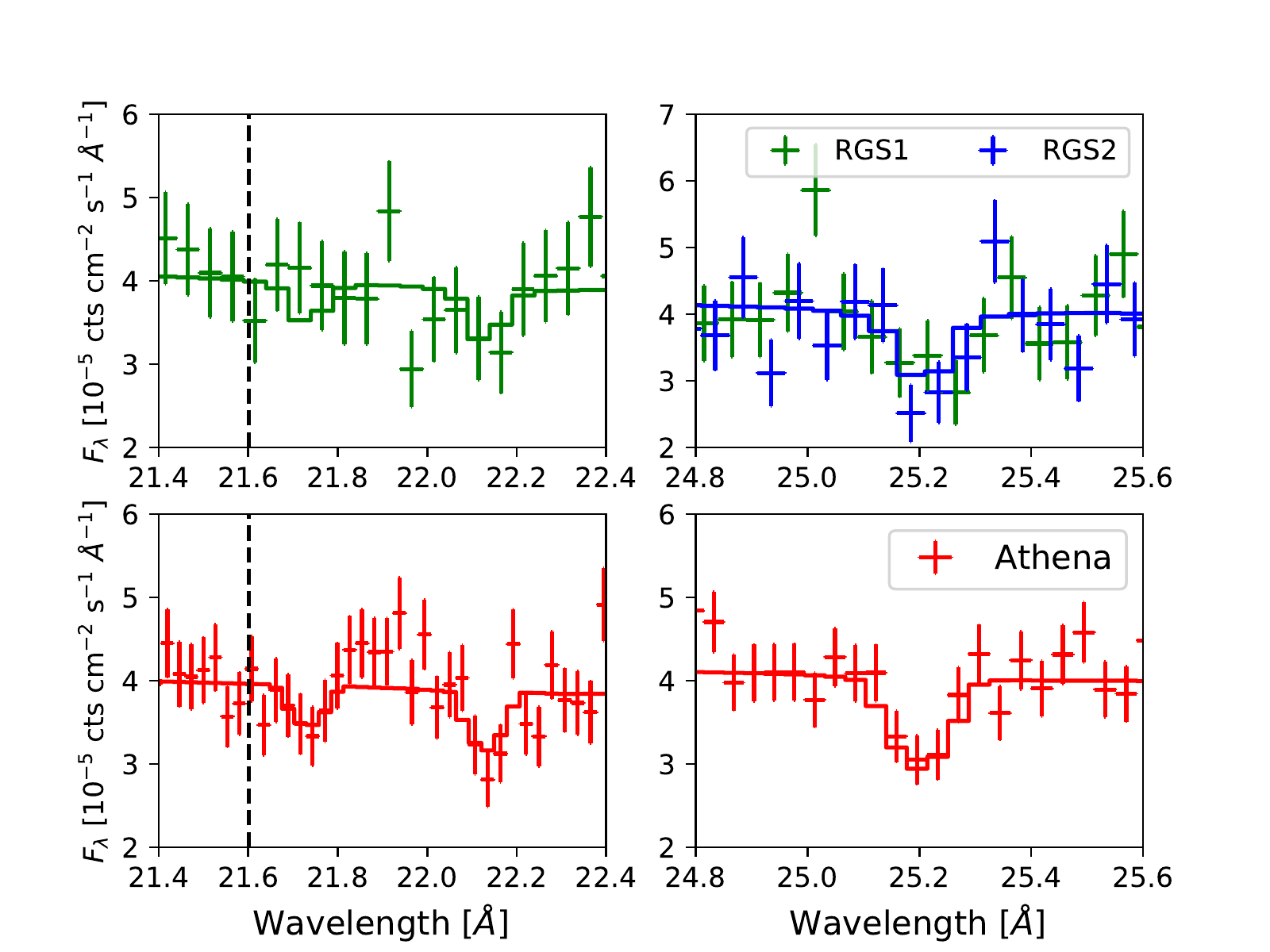}
\end{center}
\caption{The simulated 815\,ks \xmm spectra (top) and 38\,ks \textit{Athena} spectra (bottom), assuming the best-fitted \chandra model. \xmm RGS1, RGS2 and Athena IFU spectra and the models are shown with green, blue and red points and curves respectively.  The right panels show the redshifted \ovii K$\alpha$ line. The line would be detectable at $\geq6\sigma$ with \xmm  and at $\geq10\sigma$ with \textit{Athena}. The left panels show the redshifted \oviii K$\alpha$ and \ovii K$\beta$ lines.  The \oviii K$\alpha$ would be detectable at $\geq3\sigma$ with \xmmn. \oviii K$\alpha$ and \ovii K$\beta$ would be  detectable at $>5\sigma$ and $>3\sigma$ with \textit{Athena}. The vertical dashed line is the location of the $z=0$ \ovii K$\alpha$ line, which is not detected in the \chandra spectrum.}
%The dashed vertical line is at the position of $z=0$ \ovii K$\alpha$, which is not detected in the current observation, thereby reducing the confusion with the $z=0.167$ \ovii K$\beta$ line.
\label{fig:xmmsimOVII}
\end{figure}

Given the low significance of the \ovii line detection from our pilot observations with \chandran, 
it is vital to obtain model-independent results with high S/N spectra in which individual lines are detected unambiguously. 
With this goal, we simulated \xmm RGS spectra assuming the best-fit parameters from the \chandra data. In addition to detecting the \ovii $K\alpha$ line with confidence, it would be important to detect the \oviii $K\alpha$ line; the \ovii to \oviii ratio would determine the gas temperature in a model-independent way.  To detect the \oviii line at \gax$3\sigma$, $815$ ks of \xmm time would be necessary. With this exposure time the \ovii $K\alpha$ line would be detected at \gax$7\sigma$ (figure \ref{fig:xmmsimOVII}). To detect the \ovii K$\beta$ line at $\approx 3\sigma$ would require more than double this time of 1.75Ms.

%a high S/N spectrum of the \pks sightline is clearly required for detecting the warm-hot CGM unequivocally. 

Future JAXA/NASA mission XRISM will have a soft X-ray spectrometer with 5--7 eV resolution calorimeters. XRISM is expected to launch in 2021. Arcus is a high resolution X-ray grating spectrometer mission that was proposed to NASA as a medium class mission and it is selected for a Phase A concept study. The combination of high resolution and high effective area of these missions will open a new window on the CGM science. 

{\textit Athena} is the planned large X-ray astronomy mission of the European Space Agency, scheduled to launch in 2031; it is being designed to study the ``hot and energetic Universe''\footnote{https://sci.esa.int/web/athena/home}. The X-ray Integral Field Unit (XIFU) on board {\textit Athena} has high spectral resolution and large area, making it  well-suited for the studies of the warm-hot intergalactic and circumgalactic medium. We simulated  {\textit Athena} XIFU spectra of the \pks sightline assuming the best-fit parameters from the \chandra observations. With just 38ks of {\textit Athena} observations, we will detect the \ovii $K\alpha$ line at $>10\sigma$ and the \oviii line at $>5\sigma$ (figure \ref{fig:xmmsimOVII}). The \ovii $K\beta$ line will also be detected at $5\sigma$. The \ovii $K\alpha$ to $K\beta$ ratio is informative of line saturation, allowing precise determination of the column density and the velocity dispersion parameter in a model-independent way. We can constrain the nonthermal microturbulent velocity down to $20$ km s$^{-1}$ at a $3\sigma$ level. Thus the CGM science will be revolutionized with {\textit Athena}.

\section{Conclusion}

We present 376 ks \chandra LETG observations of a carefully selected galaxy in which the presence of broad \ovi together with the non-detection of \lya was indicative of warm-hot gas. Our best-fit absorber model has $\log \rm (T/K)=6.3\pm 0.2$ and $\log \rm (N_{H}/cm^{-2)}=20.7^{+0.3}_{-0.5}$. This is consistent with the warm-hot plasma model (S4) based on UV observations of SA10. The cooler models of SA10 (S1, S2, and S3) are ruled out at $99$\% confidence. However, the strongest absorption line of \ovii  is barely detected below $2\sigma$ significance in the \chandra spectrum. Thus we have a suggestive, but not conclusive evidence for broad \ovi absorber probing warm-hot gas from the shallow observations of this pilot program. 811 ks \xmm observations will clearly detect \ovii and \oviii lines, allowing us to determine the physical conditions in the CGM in a model-independent way. With just 38ks of {\textit Athena} time, we will additionally detect the \ovii $K\beta$ line, opening a new window into the CGM science.

\section*{Acknowledgments}
The scientific results reported in this article are based on observations made by the Chandra X-ray Observatory (ObsIDs: 21387, 21388, 21389, and 21955). This research has made use of software provided by the Chandra X-ray Center (CXC) in the application packages CIAO, ChIPS, and Sherpa. Support for this work was provided by the National Aeronautics and Space Administration through Chandra Award Number GO9-20121X to AG issued by the Chandra X-ray Center, which is operated by the Smithsonian Astrophysical Observatory for and on behalf of the National Aeronautics Space Administration under contract NAS8-03060. 
SM and SD gratefully acknowledges the support from the NASA grant NNX16AF49G.  YK acknowledges support from grant DGAPA-PAPIIT 106518, and from program DGAPA-PASPA. 

\vspace{5mm}
\noindent
\textit{Facilities:} \chandra, \xmmn. \\
\textit{Software:} CIAO \citep{Fruscione2006}, HeaSoft v6.17 \citep{Drake2005}, DS9 \citep{Joye2003}

\clearpage
\bibliography{reference.bib}

\clearpage

\begin{table}
\scriptsize
\caption{Best fit \chandra model and the UV models}
\begin{tabular}{lcccccc}
    &       & & &    &   &  \\
\hline
Model &  $\log \rm T$ & $\log \rm N_{H}$ & $v$ & Photon Index & Normalization & $\chi^2$  \\
        &   K   &  cm$^{-2}$ & km s$^{-1}$  & $\Gamma$ &    $10^{-4}$  & (738 dof) \\
\hline
 X-ray Best-fit:  &       & & &    &   &  \\
Continuum  &       & & &  $1.97\pm 0.04$  &  $6.9 \pm 0.1 $ &  \\
hPHASE  &   $6.3\pm 0.2$    & $ 20.7^{+0.3}_{-0.5}$ & $12^{+125}_{-12}$ &    &   &  588.28  \\
\hline 
 UV models:   &       & & &    &   &  \\
 S1   &   5.8    & 19.11 & 45.3 &    &   & 603.10 \\
 S2   &   6.0    & 19.71 & 40.8 &    &   & 600.54 \\  
 S3   &   6.11  & 19.87 & 36.8 &    &   & 598.95 \\  
 S4   &   6.41  & 20.50 & 0      &    &   & 591.39 \\
\hline
\end{tabular}
\label{tab:Allmodels}

\noindent
The power-law normalization is in the units of photons cm$^-2$ s$^{-1}$ keV$^{-1}$.\\
The UV models are from SA10.
\end{table}

\clearpage

\end{document}